\documentclass[aps, prl, showpacs, superscriptaddress, twocolumn]{revtex4}

\usepackage{amsmath}    
\usepackage{graphicx}   
\usepackage{verbatim}   
\usepackage{color}      
\usepackage[colorlinks=true, linkcolor=blue, citecolor=blue, urlcolor=blue]{hyperref}   

\usepackage{siunitx}
\usepackage{natbib}
\usepackage[final]{changes}
\colorlet{Changes@Color}{red}

\newcommand{\bs}[1]{\ensuremath{\boldsymbol{#1}}}
\newcommand{\dd}{\, \mathrm{d}}

\def\a{\beta} 

\def\gc{G_c} 

\def\Id{\mathsf{I}}
\def\kth{\mathsf{k_c}} 
\def\lbahr{\ell_0} 
\def\lnum{\ell} 
\def\Sc{\mathcal{S}}
\def\sic{\mathsf{\sigma_{\!c}}} 
 
\def\u{\mathbf{u}} 
\def\ve{\bs{\varepsilon}} 
\def\veth{\ve^{\mbox{\scriptsize \rm th}}} 
\def\x{\mathbf{x}} 

\graphicspath{{.}{figures/}}

\begin{document}
\title{Morphogenesis and  propagation of complex cracks induced by thermal shocks}
\author{Blaise Bourdin}
\thanks{Corresponding author}
\affiliation{Department of Mathematics and Center for Computation $\&$ Technology,  Louisiana State University, Baton Rouge, LA 70803, USA}
\author{Jean-Jacques Marigo} 
\affiliation{Laboratoire de M\'ecanique des Solides (UMR-CNRS 7649), \'Ecole Polytechnique, 91128 Palaiseau Cedex}
\author{Corrado Maurini}
\affiliation{Institut Jean Le Rond d'Alembert  (UMR-CNRS 7190), Universit\'e~Pierre et Marie Curie,  4 place Jussieu, 75252 Paris, France}
\affiliation{Institut Jean Le Rond d'Alembert  (UMR-CNRS 7190), CNRS, 4 place Jussieu, 75252 Paris, France}
\author{Paul Sicsic} 
\affiliation{Laboratoire de M\'ecanique des Solides (UMR-CNRS 7649), \'Ecole Polytechnique, 91128 Palaiseau Cedex}
\affiliation{Lafarge Centre de Recherche,  95 Rue de Montmurier 38290 St-Quentin-Fallavier, France }

\date{\today}

\begin{abstract}

We study the genesis and the selective propagation of complex crack networks induced by thermal shock or drying of brittle materials.
We use a quasi-static gradient damage model to perform large scale numerical simulations showing that the propagation of fully developed cracks follows Griffith criterion and depends only on the fracture toughness, while crack morphogenesis is driven by the material's internal length. 
Our numerical simulations feature networks of parallel cracks and selective arrest in two dimensions and hexagonal columnar joints in three dimensions, without any hypotheses on cracks geometry and are in good agreement with available experimental results.
\end{abstract}


\maketitle

Complex crack patterns are ubiquitous in nature and in technology applications.
Yet the theoretical understanding and predictive numerical simulation of how and when complex crack patterns arise (nucleation) and how they evolve (crack propagation) is fraught with challenges.
Although approaches based on phase fields~\cite{PonsKarmaNat2010} or variational regularizations~\cite{BouFraMar08} have led to significant advance in the numerical simulation of complex crack patterns,
short of introducing initial flaws at the structural scale~\cite{BahFisWei86}, prescribing ad-hoc stress criteria~\cite{Jag02}, or accepting global energy minimization arguments whose physical relevance is debated~\cite{Jen05,MauBouGau12,Mar10}, the predictive understanding of crack nucleation is still an elusive goal.

It is well-accepted that while Griffith--like models are appropriate for crack propagation at the scale of a structure, they are inadequate for the modeling of crack nucleation in brittle materials.
Arguably, finer models, where  a microscopic (\emph{material}) length scale plays a fundamental role, are necessary to  determine the critical load and crack geometry at the onset, especially in situations where complex crack patterns arise straight from the nucleation.
The consistent combined  modeling and numerical simulation of crack nucleation and propagation from the material to the structural  length-scale is a challenging and largely open issue.
\begin{figure}[ht]
  \centering
  \includegraphics[width=\columnwidth]{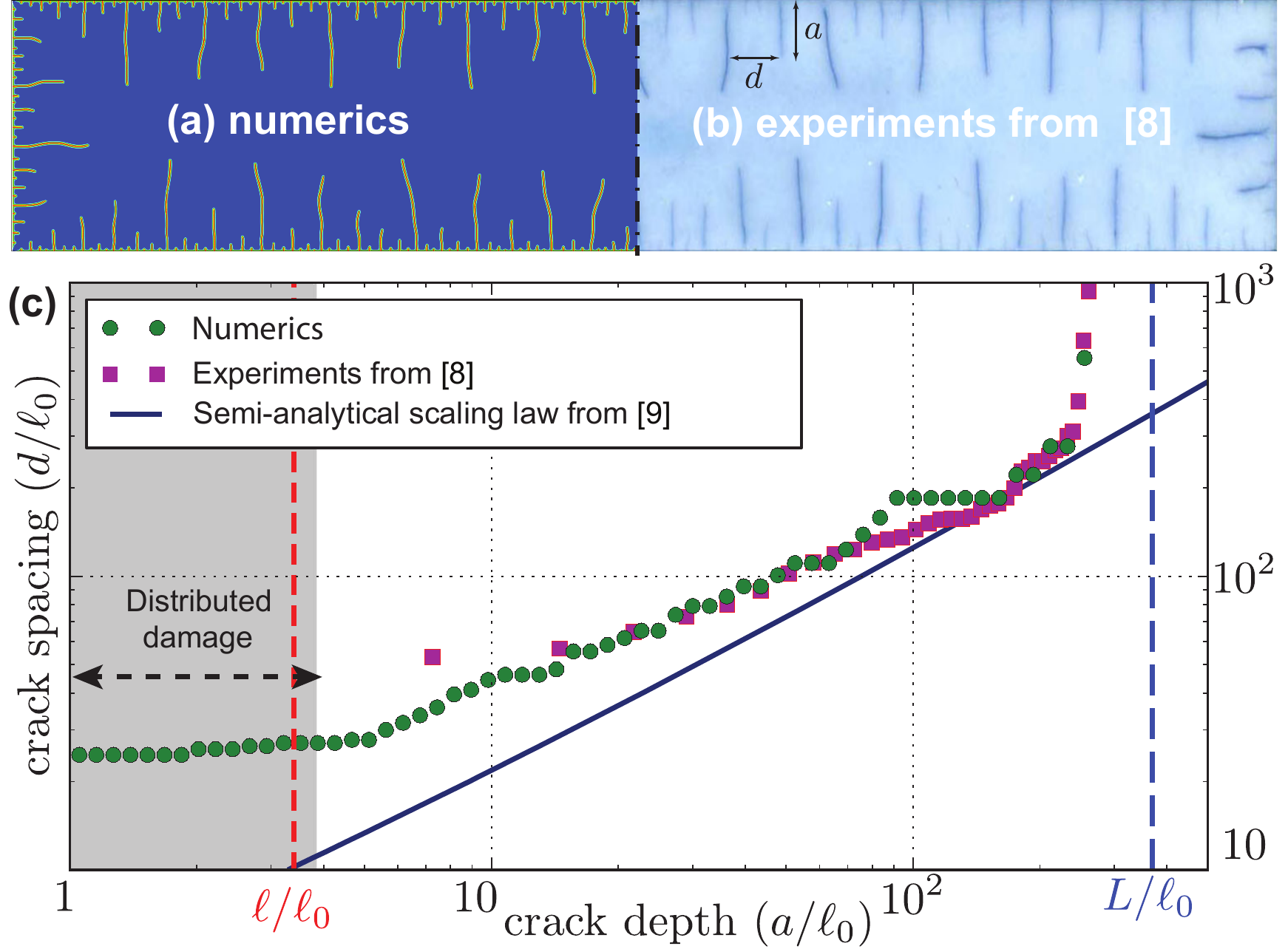}
  \caption{
  (color online). Full scale numerical simulation of a ceramic slab submitted to a thermal shock. 
  (a) Damage field from the numerical simulation (blue $\alpha=0$, red $\alpha=1$). 
  (b) Experimental results from  ~\cite[FIG. 5(d)]{ShaZhaXu11}. 
  (c) Average crack spacing  $d$ as a function of their depth $a$ for (a) and (b). 
  The solid line is an approximate scaling law obtained  in~\cite{BahWeiBah10} by imposing a period doubling condition on a Griffith model.  
  Here $\lnum=\SI{46}{\micro\metre}$ is the material internal length, $\ell_0=\gc/\left(E \a^2 \Delta T^{2}\right)=\SI{14}{\micro\metre}$ the Griffith length (loading parameter), $2\,L=\SI{9.8}{\milli\metre}$ the total depth of the slab. See FIG.~\ref{fig:nucleation} for the meaning of the distributed damage zone.
  }
  \label{fig:Shao}
\end{figure}

In this Letter, we study the morphogenesis and the selective growth of complex crack patterns induced by material shrinking under thermal shock. 
We report unprecedented quantitative agreement between numerical simulations, a theoretical model, and experiments at scales spanning from the material internal length to the structural length-scale.
Our numerical simulations predict key features of fracture patterns observed in experiments, such as the formation of periodic patterns and  the scaling laws governing their selective propagation in two  and three dimensions, and do not require any a priori hypotheses on cracks geometry.
The method we use leverages recent progress in the understanding of the links between  damage models~\cite{PhaMar10b,PhaMar12b} and the variational approach to fracture~\cite{FraMar98,BouFraMar08}. 
It is based on a rate-independent gradient damage model with stress-softening  based on two material parameters: the fracture toughness which rules the evolution of fully developed cracks, and the material's internal length which controls the initial stages of crack nucleation. \\

We investigate the thermal shock of a brittle ceramics, a now classical experimental setup~\cite{BahFisWei86,JiaWuLi12,GeyNem82} where a sample initially at a uniform temperature $T_0$ is quenched in a cold bath at temperature $T_0-\Delta T$.
We consider a rectangular slab $\Omega$ exposed to the thermal shock through its thin faces only.
We focus first on very thin slabs, which we represent by a two--dimensional body in plane stress.
We assume that within the range of temperatures involved, the material properties remain constant. Denoting by  $\u$  the displacement field  and $\ve =(\nabla \u+\nabla^T \u)/2$ the linear strain tensor,  we consider  for the sound material a linear elastic behavior of energy density $\psi_t(\ve)=A_0(\ve-\veth_t)\cdot(\ve-\veth_t)/2$ where $A_0$ is the isotropic elastic stiffness tensor.
The inelastic deformation induced by the time-dependent temperature field $T_t$ is  $\veth_t = \a (T_t-T_0) \Id$, where $\Id$ is the identity matrix.
The index $t$ is meant to highlight the dependence on time.
We neglect the cracks influence on heat transfer so that the temperature field $T_t$ solves the heat equation $\partial_t T_t - \kth\nabla^2 T_t = 0$ on $\Omega$. \replaced{Phase changes, non-uniform convection and other non-linear aspects of the heat exchange between the fluid and the sample are neglected by assuming that the temperature of the domain boundary exposed to the thermal shock is constant and equal to that of the water bath, \emph{i.e.}  $T_t=T_0-\Delta T$. }{with Dirichlet boundary conditions $T_t=T_0-\Delta T$. }
{Inertial effects are not considered because the  diffusion velocity of the temperature field is much slower than the wave speed in the material at the relevant scales in time and space. This hypothesis is universally accepted in the literature on thermal shock problems~\cite{GeyNem82,BahFisWei86,Boeck1999,Jag02,BahWeiBah10}.}
{ We} model material failure using a gradient damage model characterized by the energy function 
\begin{equation}
  \mathcal{E}_t(\u,\alpha)= \int_{\Omega} \frac{  \psi_t(\ve) }{s(\alpha)} 
+  \frac{G_c}{{4}c_w}\left(\frac{w(\alpha)}{\lnum}  + \lnum \,|\nabla \alpha|^{2} \right)\dd \x,
\label{RegularizedFractureEnergy}
\end{equation}
where $\alpha$ is a scalar damage field varying between $0$ (sound material) and $1$ (fully damaged material), $G_c$ is the material's fracture toughness,  $\ell$ an internal length, and $c_w=\int_0^1\sqrt{w(s)}\,d s$ a normalization constant. 
In a time-discrete setting, the quasi-static evolution is obtained by solving at time $t_i$   the following minimization problem 
$  \min_{\u, \alpha\ge \alpha_{i-1}} \mathcal{E}_{t_i}(\u,\alpha)$,
where the unilateral constraint on $\alpha$ enforces the irreversibility condition on the damage. 
The compliance function $s$ and the energy dissipation function $w$ should be chosen such that \eqref{RegularizedFractureEnergy} converges as $\ell \to 0$ to a Griffith--like energy $\int_{\Omega\setminus\Gamma} \psi_t(\ve) \dd \x + G_{c} \Sc(\Gamma)$, where $\Sc$ is the surface measure of the crack $\Gamma$ 
~\cite{BouFraMar00,BouFraMar08,Bra98}.
{In this model, material interpenetration in the fully damaged area is possible. In all the simulations presented here, it can be checked a posteriori that this issue does not present itself.} 
Here, we use $s(\alpha) = 1/(1-\alpha)^2$ and $w(\alpha) = \alpha$, a choice motivated by the convenience of its numerical implementation and specific analytical studies~\cite{PhaMar12b,smText}. With this choice the damage model has a stress-softening behavior and remains purely elastic without damage until the stress reaches the critical value:
\begin{equation}
\sic:=\sqrt{ \dfrac{G_c E \,w'(0)}{2 \,c_w \,\lnum\, s'(0)}}=
\sqrt{ \dfrac{3G_c E}{8\lnum }}.
\label{sigmac}
\end{equation}
The relation above may be used to determine the numerical value of the internal length for a specific material from the knowledge of its elastic limit $\sic$, Young modulus $E$, and fracture toughness $G_c$ \cite{smText}. 
The present model is in many aspects similar to the phase-field models of fracture developed independently~\cite{KarKesLev01}. Those with single--well dissipation potentials~\cite{HakKar05,PonsKarmaNat2010} are in the form of~\eqref{RegularizedFractureEnergy} with $w(\alpha)= c(1-g(1-\alpha))$, where $g(\phi)=4\phi^3-\phi^4$. 
One significant difference is that while phase-field models typically involve some form of viscous regularizations, our formulation is rate-independent. 
In addition, the current literature based on phase-field models is concerned only with the propagation of a pre-existing cracks and does not consider the initiation problem.

The dimensional analysis of the energy \eqref{RegularizedFractureEnergy} highlights three characteristic lengths: the  geometric dimension of the domain $L$, the internal length $\lnum$ and the Griffith length $\lbahr = \gc/\left(E \a^2 \Delta T^{2}\right)$. 
Using the material's internal length as the reference unit, the problem can be reformulated in terms of two dimensionless parameters, the dimension of the structure $L/\lnum$ (a geometric parameter) and  the intensity of the thermal shock $\lbahr/\lnum$ (a loading parameter).
This is a significant departure from the classical Griffith setting where the only relevant parameter is $L/\lbahr$ \cite{Jag02,Jen05,BahWeiBah10}.\\

Figure~\ref{fig:Shao} compares the experiment from~\cite[FIG.~5(d)]{ShaZhaXu11} ($\SI{1}{\milli\metre} \times \SI{9.8}{\milli\metre} \times \SI{50}{\milli\metre}$ ceramic slab, $\Delta T = \SI{380}{\degreeCelsius}$) with the damage field from a numerical solution of the gradient damage model. 
The material properties, communicated by the authors of~\cite{ShaZhaXu11} are $E=\SI{340}{\giga\pascal}$, $\nu = 0.22$, $G_c = \SI{42.47}{\joule\meter^{{-2}}}$, $\sigma_c=\SI{342.2}{\mega\pascal}$, and $\beta = \SI{8e-6}{K^{-1}}$, which using \eqref{sigmac} gives $\lnum=\SI{46}{\micro\metre}$ and $\ell_0=\SI{14}{\micro\metre}$.
{As our model is rate independent its solution are independent of $\kth$, up to a change of time scale.}
The numerical results are obtained through a finite element discretization and the approach of~\cite{Bou07a,BouFraMar08,smText}. 
The main technical difficulties are the constrained minimization of a non-convex energy, and the need for a spatial discretization adapted to the material length-scale scale $\lnum$.
Cracks correspond to the localized bands where $\alpha$ goes from 0 to 1 and back to 0. 
The qualitative agreement between experiments and simulation is very good. 
In particular, our simulations reproduce the key phenomenon: the emergence of a periodic array of parallel short cracks at the initiation and their selective propagation toward the interior of the slab.
Figure~\ref{fig:Shao}(c) shows a quantitative comparison between the numerical simulation of FIG.~\ref{fig:Shao}(a) and experimental data from~\cite{ShaZhaXu11} by plotting the average crack spacing $d$ as a function of the distance $a$ to the edge exposed to the thermal shock for the final configuration, the agreement is striking. 
Note that in the experimental results shorter cracks are probably filtered out by the adopted experimental crack detection methods~\cite{smText}. 
In a first regime, very short equi-distributed cracks nucleate (the plateaus of the crack spacing for short depth in the numerical experiments), followed by selective arrest and period doubling  \cite{video}.
In the central region of the plot, we can compare experimental and simulation data with a scaling law obtained in~\cite{BahWeiBah10} through linear fracture mechanics calculations by imposing a bifurcation condition between crack propagation modes with period doubling or not (solid line). 
For larger values of $a$, we observe the final crack arrest caused by the finite size of the sample, again in very good agreement with the experiments.
Whereas classical theories can be applied in the second and third regimes consisting of fully developed cracks, they cannot properly account for the nucleation phenomenon observed here \added{without preexisting flaws. Our simulations are initialized with a null damage field, an homogenous material, and an unflawed geometry. The crack nucleation is due to the softening character of the material behavior.}\\

\begin{figure}[t]
  \centering
   \includegraphics[width=\columnwidth]{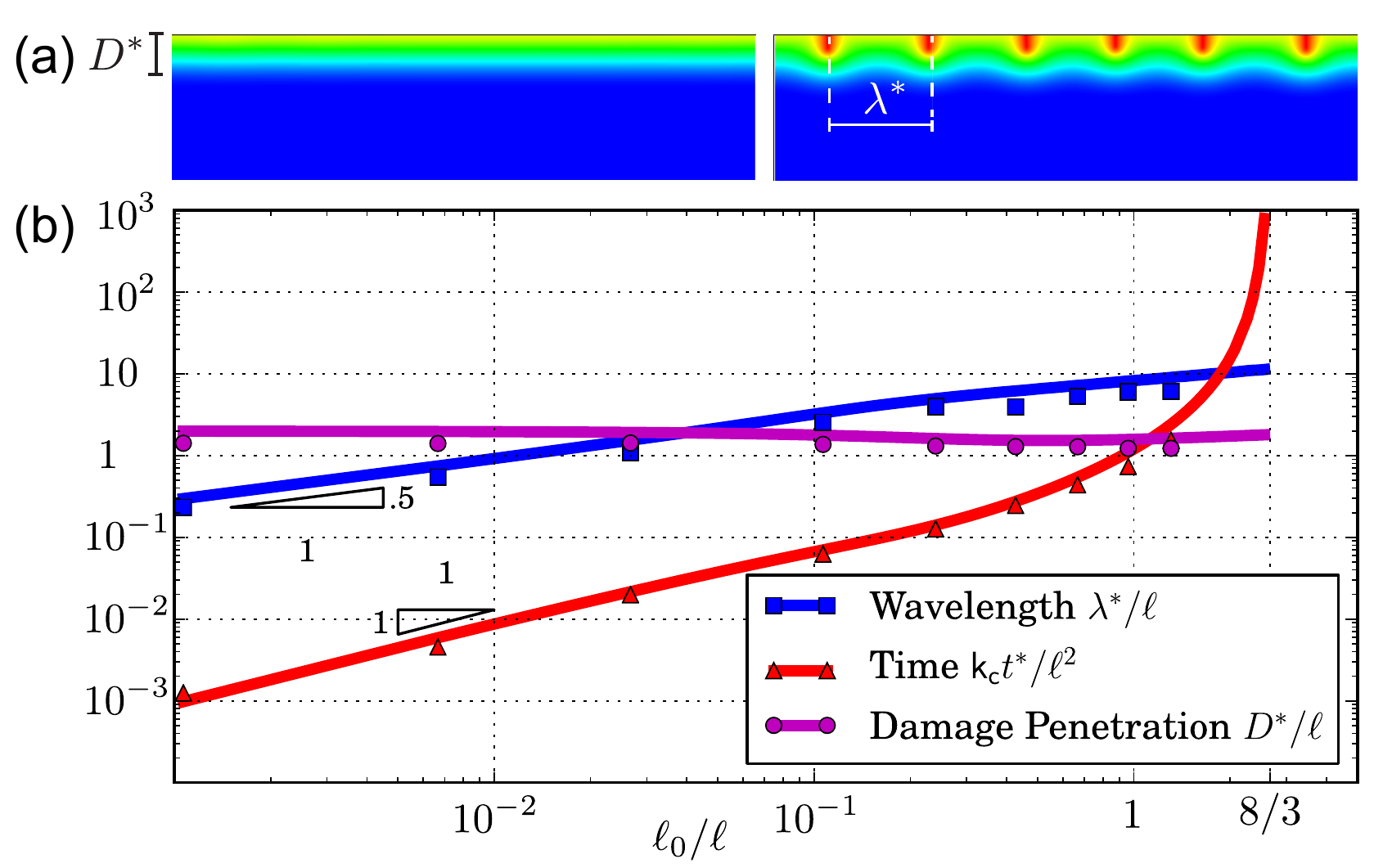}
  \caption{(color online).
Crack nucleation: 
(a) Damage field near the shock surface before (left) and after (right) bifurcation time $t^*$ for $\lbahr/\lnum=0.107$ showing the bifurcation of a horizontally-homogeneous damaged band of depth $D^*$ toward a periodic solution with wavelength $\lambda^*$.
(b) Wavelength, time and damage penetration in numerical simulations for several intensities of the thermal shock  $\lbahr/\lnum$ (dots), compared to the semi-analytical results from~\cite{SicMarMau13}  (solid lines).
}
  \label{fig:nucleation}
\end{figure}

The second series of  simulations focuses precisely on the crack nucleation process and hence on short times.
In this setting, one can assume that the domain is semi-infinite so that the geometric parameter $L/\lnum$ is infinite and the only parameter is the intensity of the thermal shock $\lbahr/\lnum$.
For an undamaged material, the stress is uniaxial and reaches its maximum value $\sigma_\mathrm{max} = E \beta \Delta T$ at the surface of the thermal shock. 
Since $(\sigma_\mathrm{max}/\sigma_c)^2  =  3\lnum/8\lbahr$, for mild-enough thermal shocks ($\lbahr/\lnum>8/3$), the critical stress is never reached and the solution remains elastic at all time. 
If $\lbahr/\lnum<8/3$, damage takes place at $t=0$, is homogeneous in the horizontal direction, and non-null in a band  of finite thickness $D$, which penetrates progressively inside the body until a critical time $t^*$.
At $t=t^*$ the horizontally--homogeneous solution becomes unstable and the damage field develops oscillations of periodicity $\lambda^*$ (FIG.~\ref{fig:nucleation}(a)).
{
An analytical solution for the damage field in the first stage of the evolution and its bifurcation and stability analysis is  reported in \cite{SicMarMau13}, providing semi-analytical results for the periodicity $\lambda^*$, the damage penetration $D^*$, and the time $t^*$ at the bifurcation. 
Here we perform several simulations varying $\ell_0/\ell$ and detect the critical parameter at the bifurcation. In  Figure~\ref{fig:nucleation}(b)  the numerical simulations (dots) are compared to \cite{SicMarMau13} (solid lines). The good agreement provides an excellent verification of our numerical model.
}
For severe shocks ($\lbahr \ll \lnum$), the results disclose a well-definite asymptotic behavior with $\lambda^*\sim \sqrt{\lbahr \lnum}$, $D^*\sim\ell$, and $t^*\sim \lbahr \ell/\kth$.
In this regime we observe numerically that all oscillations at the bifurcation develop in fully formed cracks ($\max{\alpha}=1$), which is not the case for milder shocks ($\lbahr \sim \lnum$).
However, the full post-bifurcation analysis remains an open problem at this time.

Experimental studies show that in three-dimensions cracks delineate cells with coarsening polygonal cross-sections~\cite{GoeMahMor09}. 
Because of the complexity of the problem, the few available theoretical and numerical studies are based either on simplified two--dimensional models~\cite{JagRoj02, Jun12} or on strong assumptions on the crack geometry~\cite{BahretalPRE2009}. 
The numerical simulation and analysis of the full three--dimensional problem is a major challenge for classical fracture mechanics tools and remains therefore largely unexplored.  

\begin{figure}[th]
\centering
\includegraphics[width=\columnwidth]{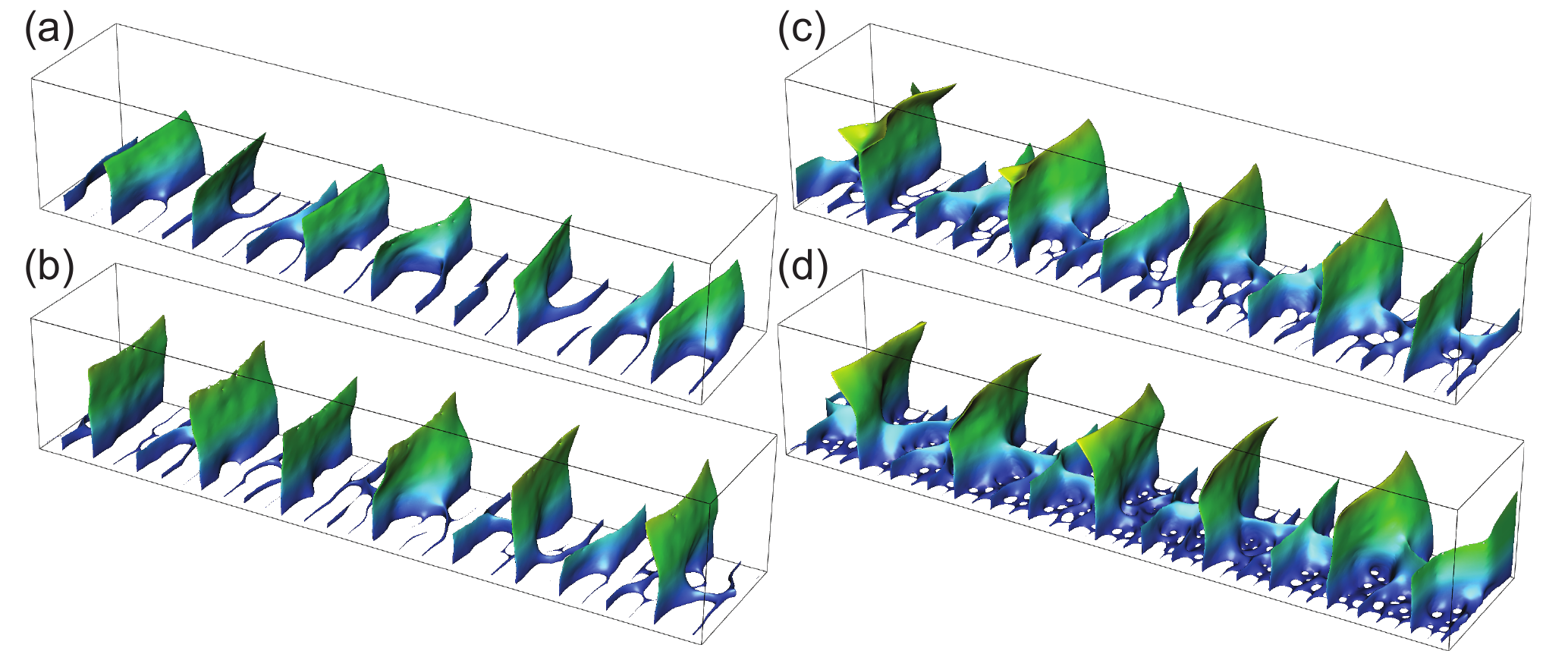}
\caption{(color online). 
	{Three--dimensional version of the experiment from FIG.~\ref{fig:Shao}(b) showing the transition from two to three--dimensional crack patterns.  
	The simulations are performed on a subdomain of dimension $\SI{5}{\milli\meter} \times \SI{1}{\milli\meter} \times\SI{1}{\milli\meter} $  and temperature contrast (a) \SI{380}{\degreeCelsius} ($\lbahr= .27\,\lnum$);  (b) \SI{480}{\degreeCelsius} ($\lbahr=0.17\,\lnum$); (c) \SI{580}{\degreeCelsius} ($\lbahr=0.12\,\lnum$); (d) \SI{680}{\degreeCelsius} ($\lbahr=0.08\,\lnum$).}
	}
\label{fig:Temp3D}
\end{figure}

 \begin{figure}[ht!]
\centering
\includegraphics[width=\columnwidth]{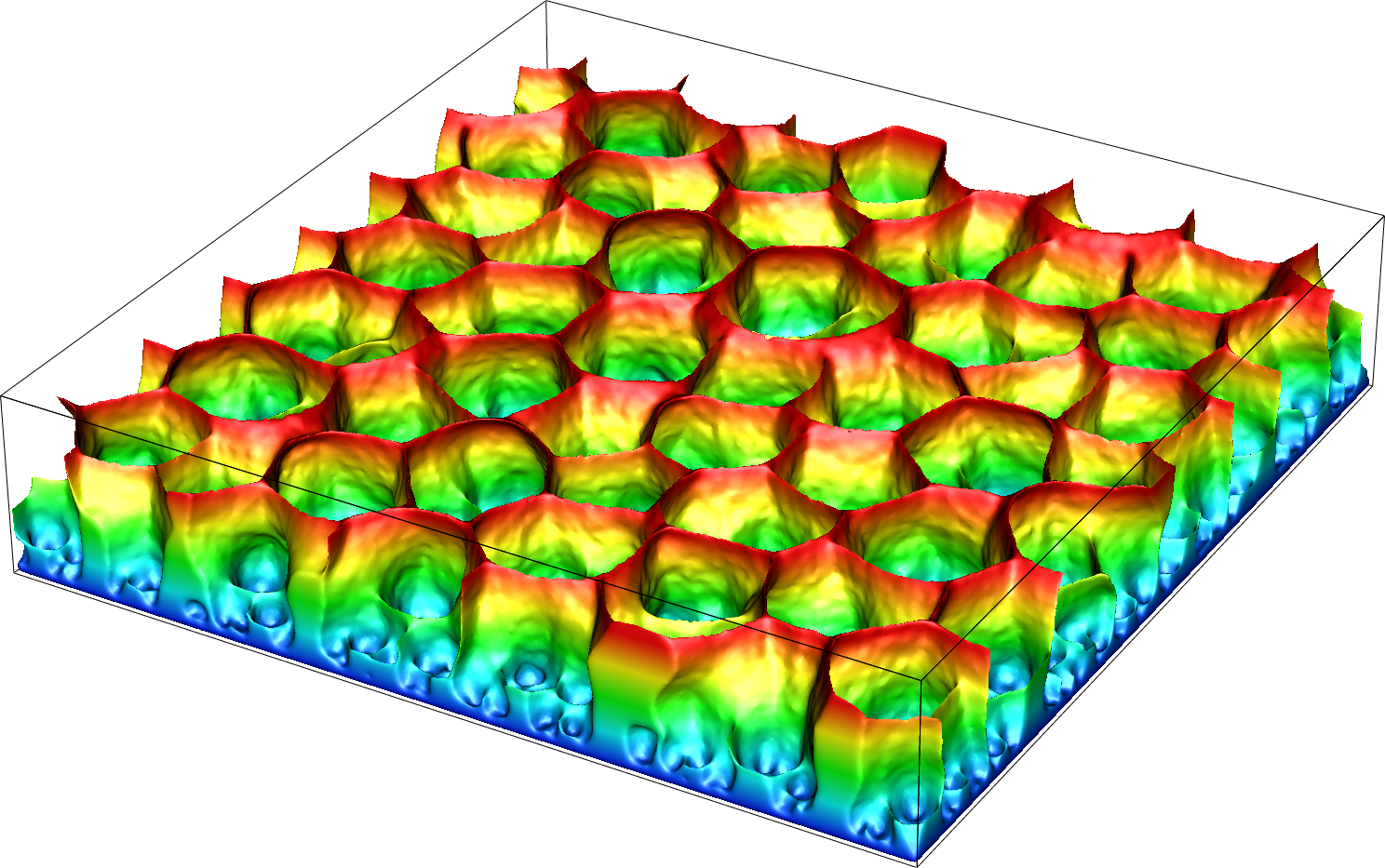}
\caption{(color online).
Complex fracture pattern  for $\ell_0=\gc/\left(E \a^2 \Delta T^{2}\right)=0.05\,\ell$ in a domain of size $150\ell  \times {150}{\ell} \times {20}{\ell}$ color-coded by distance from the bottom surface where the thermal shock is applied. 
The problem was discretized in 44M linear finite elements in space (mesh size $h=\ell/5$) and 100 time steps. 
The computation was performed on 1536 cores of the NSF-XSEDE cluster Stampede at Texas Advanced Computing Center in \SI{10}{h}.
}
\label{fig:1282061}
\end{figure}
\begin{figure}[ht!]
  \centering
  \includegraphics[width=\columnwidth]{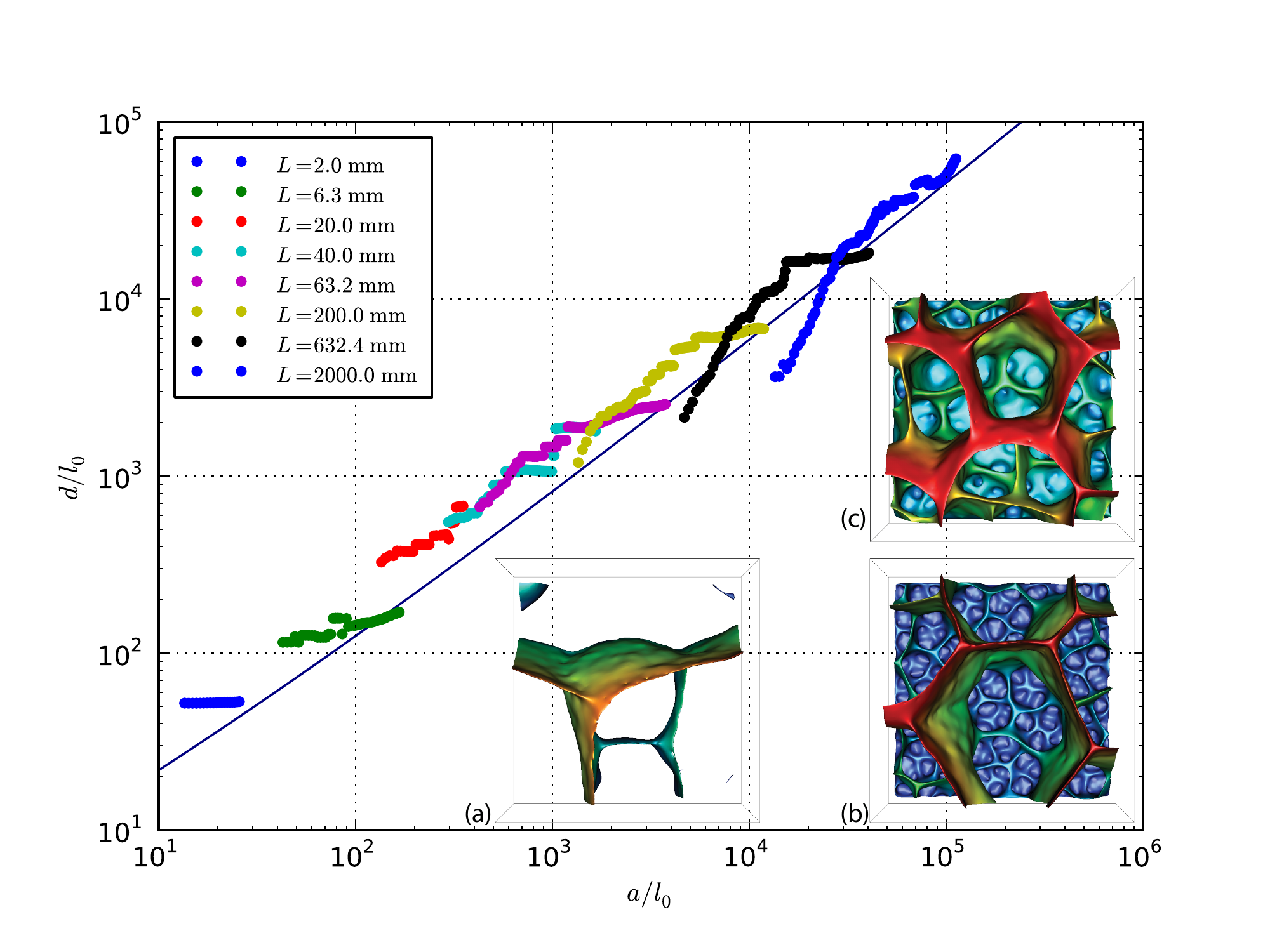}
  \caption{(color online). 
 Average size $d$ (square root of the average cross-sectional area) of the fracture--delimited cells as a function of the depth $a$ (distance to the exposed face) in cubic domains with edge length $L$ ranging from \SI{2}{\milli\meter} to \SI{2}{\meter} compared with the two--dimensional scaling law from~\cite{BahWeiBah10} (solid line). All the simulations are with $\ell_0=\SI{14}{\micro\meter}$ and $\ell=L/40$. Inset:   top view to the   crack patterns for (a) $L=\SI{2}{\milli\meter}$; (b) $L=\SI{63.2}{\milli\meter}$; (c) $L=\SI{2}{\meter}$.}
  \label{fig:Scaling3D}
\end{figure}

Figure~\ref{fig:Temp3D} is a three--dimensional version of the simulation from FIG.~\ref{fig:Shao} on a plate of thickness  
\SI{1}{\milli\meter}, for increasing values of $\Delta T$.
The fracture geometry is represented by the level surface $\alpha = 0.95$.
In order to reduce the computational cost, the computation is performed on a  fragment of width \SI{5}{\milli\meter} and height \SI{1}{\milli\meter} of the domain, and the temperature is assumed constant throughout the thickness of the sample. 
We observe a transition from transverse cracks to three--dimensional fracture patterns delimiting polygonal cells between \SI{480}{\degreeCelsius} and \SI{580}{\degreeCelsius}, which is consistent with~\cite[Fig~5]{ShaZhaXu11}. 
Another series of simulations performed at constant temperature for increasing sample thickness (not shown here) highlight the same behavior: transverse cracks for thin domains, transitioning to three--dimensional cracks for thicknesses between $\SI{1}{\milli\meter}$ and $\SI{2}{\milli\meter}$.
This is also consistent with the observations in~\cite[Fig~5]{ShaZhaXu11} and justifies the use of a two--dimensional model, a posteriori.

Figure \ref{fig:1282061} shows a fully tri--dimensional crack pattern obtained for a domain of dimension $150\ell  \times {150}{\ell} \times {20}{\ell}$ for $\ell_0=0.05\,\ell$.
During the simulation, a  disordered pattern of small cells nucleates in the first time steps and propagates quasi-statically inside the domain.
A selection mechanism leading to honeycomb patterns with increasingly large and regular cell arises from energy minimization.
\added{Tracking the propagation of three-dimensional crack front of Figure~\ref{fig:1282061} using a classical Griffith--based model requiring an explicit description of the crack surface and its propagation criterion would be prohibitively complex. Instead, our three-dimensional computations are performed through a straightforward extension of the  discretization and minimization  algorithm  for the energy~\eqref{RegularizedFractureEnergy}. }
Obtaining an accurate scaling law for the cell diameter as a function of the depth as the one of the two--dimensional case of Figure \ref{fig:Shao} would require simulations on larger domains which would rapidly become computationally prohibitive. 
Hence, we perform a series of numerical experiments by fixing the loading $\ell_0$ and by varying at the same time the internal length $\ell$  and the domain size $L$ so as to keep their ratio equal to $L/\ell=40$. 
The number of element is kept constant with a mesh size $h=\ell/5$. 
For each computation, we compute  the average cell diameter $d$ as a function of distance from the bottom edge $a$ using a post-processing software. 
This process does not involve any adjustable parameter, yet our results match the two--dimensional scaling law of~\cite{BahWeiBah10} over several orders of magnitude, leading us to conclude that the scale selection mechanism in two and three dimensions are identical.
In addition, while the initial phase of the evolution depends strongly on $\ell$, later time evolution of fully developed cracks at the structural is unaffected by this parameter, matching the general scaling law for a Griffith--based model.
This finding is consistent with the properties of the energy functional $\mathcal{E}_t$ which is known to lead to a Griffith-type propagation criterion~\cite{Bra98,HakKar05,HakKarJMPS2009,SicMar13}. 
\medskip

Our simulations show that a purely quasi-static model based on energy minimization can fully explain the formation of imperfect polygonal patterns and their selective coarsening as ``maturation'' mechanism during propagation, a phenomenon sometimes attributed to non-equilibrium processes~\cite{GoeMahMor09}.   
We show that a carefully chosen gradient damage model can be used to account simultaneously for the nucleation of complex crack patterns and their propagation following Griffith criterion. 
Further works will be carried on to perform a careful statistical analysis of the geometry of the 3D crack patterns and further comparisons to experimental results.  The present modeling framework has a general validity and can be applied to  other domains including for example the formation of basalt columns with uniform cross-sectional diameters through the solidification of lava fronts~\cite{GoeMahMor09} or shaping of biological systems as observed of the scales on the heads of crocodiles~\cite{Mil13}. 
{We are also considering stronger thermo-mechanical coupling including the effect of cracks on heat transfer as in~\cite{Boeck1999,Bud94}.} 

\medskip
\begin{acknowledgments}
The authors wish to thank Yingfeng Shao for  providing the  experimental data used in FIG.~\ref{fig:Shao}. B.B. work was supported in part by the National Science Foundation grant DMS-0909267. Some numerical experiments were performed using resources of the Extreme Science and Engineering Discovery Environment (XSEDE), which is supported by National Science Foundation grant number OCI-1053575 under the Resource Allocation TG-DMS060014. J.J.M. and C.M. gratefully acknowledge the funding of the ANR program T-Shock OTP J11R087.

\end{acknowledgments}

\bibliography{prl_thermal}
\end{document}